\documentclass[12pt,preprint]{aastex}
\usepackage{psfig,natbib}

\newcommand{\citepeg}[1]{\citep[{e.g.,}][]{#1}}

\def\lsim{\hbox{ \rlap{\raise 0.425ex\hbox{$<$}}\lower 0.65ex\hbox{$\sim$}}}
\def\gsim{\hbox{ \rlap{\raise 0.425ex\hbox{$>$}}\lower 0.65ex\hbox{$\sim$}}}
\def\arcmin{\hbox{$^\prime$}}
\def\arcsec{\hbox{$^{\prime\prime}$}}

\def\fh{\hbox{$~\!\!^{\rm h}$}}
\def\fm{\hbox{$~\!\!^{\rm m}$}}
\def\fs{\hbox{$~\!\!^{\rm s}$}}
\def\ale{\mathrel{\hbox{\rlap{\hbox{\lower4pt\hbox{$\sim$}}}\hbox{$<$}}}}
\def\age{\mathrel{\hbox{\rlap{\hbox{\lower4pt\hbox{$\sim$}}}\hbox{$>$}}}}

\begin{document}

\title{The First Two Host Galaxies of X-ray Flashes: 
XRF\,011030 and XRF\,020427}

\author{J. S. Bloom\altaffilmark{1,2}, D.~Fox\altaffilmark{3}, 
P.~G.~van Dokkum\altaffilmark{4}, S.~R.~Kulkarni\altaffilmark{3},
E.~Berger\altaffilmark{3}, S.~G.~Djorgovski\altaffilmark{3},
and D. A. Frail\altaffilmark{5}}

\affil{$^1$ Harvard-Smithsonian Center for Astrophysics, MC 20, 60 Garden Street, Cambridge, MA 02138, USA}

\affil{$^2$ Harvard Society of Fellows, 78 Mount Auburn Street, Cambridge, MA 02138, USA}

\affil{$^3$ Palomar Observatory 105--24, California Institute of Technology,
            Pasadena, CA 91125, USA}

\affil{$^4$ Department of Astronomy, Yale University, New Haven, CT 06520-8101, USA}

\affil{$^5$ National Radio Astronomy Observatory, Socorro, NM 87801, USA}

\begin{abstract}
Given the paucity of empirical constraints, the nature of the
newly-recognized phenomena called X-Ray Flashes (XRFs) has been an
open question. However, with the recent detections of radio and X-ray
afterglow it is finally possible to study the large- and small-scale
environments of XRFs. We present {\it Chandra}, {\it Hubble Space
Telescope} (HST), and Keck observations of the fields of XRFs 011030
and 020427. Astrometric comparisons of the X-ray transient positions
and the HST images reveal the XRFs to be associated with faint blue
galaxies. Photometric evidence of these putative hosts suggests that
these two XRFs originated from redshifts less than $z\sim$3.5, and
thus cannot be due to GRBs at very high redshifts. In both host--burst
offsets and host properties, these XRFs could have been drawn from
distributions similar to those measured of gamma-ray bursts (GRBs).
We conclude with a discussion of the implications of this XRF--GRB
host connection for the possible progenitors of XRFs.
\end{abstract}


\section{Introduction}

A new class of high energy transients called X-ray Flashes (XRFs) has
been identified (see \citealt{hzk03} and \citealt{yin03a} for recent
reviews). These events have an annual all-sky rate which is between
one third to one half of the gamma-ray burst (GRB) rate and thus make
a substantial contribution to the cosmic explosion rate. In many
respects, particularly in their duration, the {\it prompt} burst
properties of XRFs overlap with those of the class of long duration
gamma-ray bursts.  Within the statistics limited by the small number
of identified events, about 30 thus far, XRFs appear to be isotropic
and inhomogeneous, suggestive of a cosmological distribution (as with
GRBs). However, the principal difference, as connoted by the
nomenclature is that the bulk of the energy of XRFs is measured in the
X-ray band, with a peak energy $E_{\rm p}$ below roughly 50 keV
\citep{kwh+01,bol+03}. In contrast, in the extensive sample of GRBs detected
by {\em BATSE}, $E_{\rm p}$ is clustered around 200 keV with a
distinct roll-off toward lower values of $E_{\rm p}$ \citepeg{pbm+00}.

The discovery of X-ray \citep{hyf+01,acf+02} and radio \citep{tfk01}
afterglows are consistent with XRFs being of cosmological origin
\citep{hzk03}. However, whereas at least some long-duration GRBs are a 
result of the death of massive stars (see \citealt{bloom03a} for
review), the physical mechanism(s) responsible for the production of
XRFs is completely unknown.  Theoretical models for the origin of
XRFs have been explicated elsewhere
\citep{yin02,wzh02,zm02,bol+03,mdba03}.  To summarize, XRFs could arise from 
a new physical class of explosions, GRBs which originate from very
large redshifts ($z \age 6$; \citealt{hzk+01}), or lower-redshift
variants of GRBs (e.g., GRBs beamed away from Earth: \citealt{yin02},
GRBs with dense ambient gas, or transition GRBs with lower Lorentz
factor outflows: \citealt{dcb99}).  To our knowledge, the only reason
to possibly associate XRFs with the death of massive stars is that
XRFs appear to have a duration distribution similar to those of the
long-duration GRBs.

A basic discriminator of the various XRF progenitor models is a
measurement of the distance to the explosions. In the absence of a
direct redshift measurement, it is possible to constrain the distance
by examining photometric and morphological properties of the host
galaxies of XRFs. Irrespective of the redshifts, the nature of the
hosts themselves and the location of XRFs within their hosts, in
analogy with GRBs, will play an important role in understanding the
progenitors. Here we report on the host galaxies of the first two XRFs
with sub-arcsecond afterglow
localizations\footnotemark\footnotetext{At least one additional XRF
has been subsequently localized to a sub-arcsecond
position. \citet{fkc+03} discovered the apparent optical/IR afterglow
of XRF\,030723
\citep{pbc+03}. \citet{spf+02} found a possible optical transient of
XRF\,020903, which was associated with a low redshift star forming
galaxy ($z = 0.25$).}. In both cases, we identify a putative host
galaxy. Here we present accurate astrometry of the XRFs and describe
the properties of the hosts. Finally, we compare the properties of the
host galaxies of GRBs and find that the XRF host galaxies are quite
similar to those of GRB galaxies, namely typical star forming galaxies
at moderate redshifts. The discoveries of the hosts discussed herein
have been previously announced (XRF\,011030: \citealt{fpk+02};
XRF\,020427:
\citealt{ctgsf02}).

\section{Observations and Reduction}

Our X-ray observations with the {\it AXAF CCD Imaging Spectrometer}
(ACIS) on-board the {\it Chandra} X-ray Observatory (CXO) were first
reported in \citet{hyf+01} for XRF\,011030 and in
\citet{fox02} and \citet{fox02a} for XRF\,020427. The data were reduced and
analyzed using the {\it CIAO} software
package\footnotemark\footnotetext{See {\tt
http://cxc.harvard.edu/ciao/}}. The X-ray afterglow of XRF\,011030 was
identified by \citet{hyf+01} in a 47 ksec exposure beginning on 2001
Nov 9.73 UT, consistent with the radio transient position
\citep{tfk01}. A second 20 ksec epoch was obtained on 2001 Nov 29.44
UT. The sub-arcsecond location of the X-ray afterglow of XRF\,020427
\citep{acf+02} was identified in the CXO imaging as a fading point
source between two CXO pointings (beginning 2002 May 6.24 UT: 13.8
ksec and 2002 May 14.19 UT: 12.5 ksec). The fiducial absolute
positions of the respective afterglow and field sources surrounding
the XRFs were found using {\it CIAO}/Wavdetect.

For optical/IR imaging, the field of XRF\,011030 \citep{zhl+01} was
observed starting on 12.19 December 2001 UT as part of the Cycle 9
{\it HST} observing program GO 8588 (A.~Fruchter, P.I.; see
\citealt{fpk+02}).  The field of XRF\,020427 \citep{zrg+02} was
observed starting on 14.62 June 2002 UT as part of our large Cycle 10
project GO 9180 (S.~Kulkarni, P.I.). Both fields were observed in the
STIS/50CCD (``Clear'') and STIS/F28$\times$50LP (``Longpass'')
filters. The image frames were retrieved from the STSCI archive after
``On the Fly'' preprocessing, where the raw data are pre-reduced using
the best calibration data available at the time of
retrieval. Individual exposures ranged from integrations of 864--1008
sec (XRF\,011030) to 572--624 sec (XRF\,020427). The total integration
times were 8640 s and 9072 s (XRF\,011030) and 4781 s and 4796 s
(XRF\,020427), for the Clear and Longpass filters respectively.

We combined the exposures and removed cosmic-rays using the standard
methods outlined in IRAF/DRIZZLE2 \citep{fh97}. Since few sources were
detected in the Longpass images of XRF\,020427, extra care was taken
to remove cosmic rays before cross-correlating the exposures to find
the relative offsets. In particular, we ran (for both filters) a
Laplacian detection algorithm ({\sc LACOSMIC}; \citealt{vdok01}) on
the images to detect and mask cosmic-rays before running the DRIZZLE
routine PRECOR. In the XRF\,011030 images, since several stars were
saturated, we masked saturated pixels before performing the
cross-correlation. The combined images were made using {\tt
drizzle.pixfrac}=0.7 and the final scale was 25.3 milliarcsec (mas)
$\times$ 25.3 mas per pixel. In the drizzle process, images were
rotated to cardinal orientation using the header information about the
roll angles. We registered the images of the field in two filters
using IRAF/CROSSCOR.

We also obtained supporting groundbased imaging of the XRF fields. For
XRF\,020427, three 600 s $I$-band exposures were taken using the {\it
Wide Field Reimaging CCD Camera} at the Las Campanas Observatory (LCO)
DuPont 100-inch telescope on 3 August 2002. After reductions, the
images were registered and stacked yielding a 25\arcmin $\times$
25\arcmin\ final image centered on the XRF afterglow position. The
effective seeing was 1.6\arcsec\ FWHM. On 24 December 2001 UT, we
obtained 4500 sec of $K_s$-band imaging of the field of XRF\,011030
using the NIRC instrument \citep{ms94} mounted on the Keck I 10 meter
telescope in Mauna Kea, Hawaii. The zero-point of the combined image
was determined from observations of four \citet{pmk+98} NIR standard
stars, with an estimated zero-point uncertainty of 0.05 mag.

\section{Astrometry: The location of the XRFs}

\subsection{XRF\,011030}

The HST STIS/Clear field includes apparent counterparts to three
sources detected in our CXO imaging.  An 11th magnitude star to the
Southwest, which is the apparent counterpart of one of the CXO
sources, is listed in the Tycho-2 catalog as TYC\,4590-00070-1
\citep[$d = 16.4
\pm 11.9$ pc, proper motion $100 \pm 181$ mas yr$^{-1}$, 
$V =11.04$\,mag;][]{hfm+00}.  The other two objects, to the northeast
in the STIS field, appear extended in the STIS/Clear image. However,
we are reasonably confident that these sources provide a good
astrometric tie since the first source appears to be a point-source
superimposed on a galaxy and the other source has a smooth surface
brightness profile and a well-defined center.

We centroided the HST and the CXO counterpart sources using a
Gaussian-weighted fit. However, the Tycho-2 source is severely
saturated in the HST image and so a direct centroiding proved
difficult. Instead, we found the position of the HST source using the
four diffraction spikes to find the intersection. The estimated error
on this centering method is 30 mas (2 $\sigma$).

Comparing the nominal CXO and STIS/Clear spacecraft positions we
require a shift of the CXO coordinates by 290 mas East and 60 mas
South for the best fit.  The uncertainty in this shift, taken as the
mean of the standard deviation of the three offsets, is 80 mas (2
$\sigma$). [Justified by the Tycho-2 measurements of the comparison
stars, we assume that the proper motion/parallax between the XRF and
HST epoch of the Tycho-2 source is significantly smaller that this
r.m.s.~scatter.]  Since the uncertainty in the CXO coordinate shift is
derived with only a few degrees of freedom, we consider this a
systematic uncertainty. The internal (Poissonian) uncertainty in the
CXO centroid of the XRF afterglow is 48 mas (2 $\sigma$). Adding the
errors (not in quadrature), the conservative uncertainty radius for
the XRF position on the STIS image is 128 mas (2 $\sigma$).

We have been able to make an independent registration of the CXO
position by using the unified source catalog from VLA observations
\citep{tfk01}.  Four CXO sources, all distinct from the sources
used for the STIS frame tie, have VLA counterparts. Correcting the CXO
coordinates to the VLA frame (itself closely tied to the International
Coordinate Reference Frame; ICRF) gives an adjustment of $230
\pm 530$ mas East, $120 \pm 400$ mas South, which is consistent with 
the HST adjustment we have made.  Source positional uncertainties for
this analysis are dominated by uncertainties in the radio centroiding,
as the VLA data were taken in D-array (elliptical beamsize of
19.8 $\times$ 16.5 arcsec).

Identification of the Tycho star TYC\,4590-00070-1 in the X-ray data,
and confirmation of the roll angle and overall distortions by
comparison with the HST and VLA data (see above) allows us to
determine the absolute ICRF position of XRF~011030 with confidence.
We find that the XRF afterglow is located at
$\alpha$(J2000)=20\fh45\fm36\fs.007 $\pm \,0\arcsec.060$,
$\delta$(J2000) = $+$78$^\circ$06\arcmin01\arcsec.09 $\pm
\,0\arcsec.066$. This $\sim$300 mas shift from the CXO position is entirely 
consistent with the absolute pointing accuracy (0.6\arcsec, 90\%
confidence) of {\it Chandra}\footnotemark\footnotetext{See {\tt
http://cxc.harvard.edu/cal/ASPECT/celmon/}.}.

\begin{figure*}[p] 
\centerline{\psfig{file=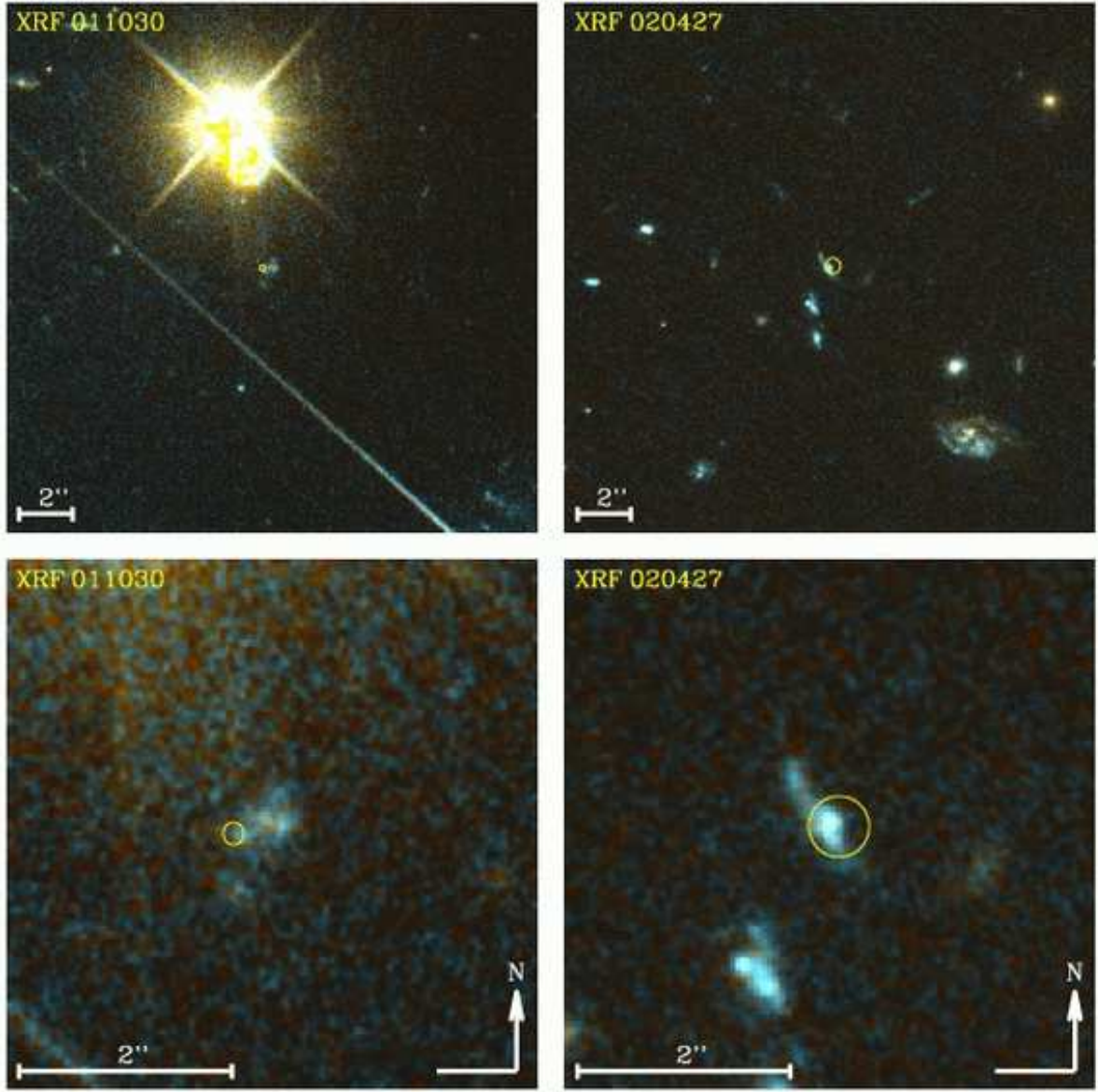,width=6.5in,angle=0}}
\caption[XRF Hosts] {The location and hosts of XRF\,011030 and
XRF\,020427 with false color images constructed from STIS Clear and
Longpass observations.  Top panels show the 20 $\times$ 20
arcsec$^{2}$ field around the XRF positions. Lower panels show the
detail of the host regions; ellipses depict the 2-$\sigma$ position of
the XRF from the {\it Chandra} localization of the afterglows. North
is up and East is to the left. 

{\bf High-resolution version at
http://www-cfa.harvard.edu/$\sim$jbloom/xrfs.pdf}.}
\label{fig:profiles}
\end{figure*}

An image depicting the position of the XRF on the HST image is shown
in Figure \ref{fig:profiles}. As noted previously \citep{fpk+02}, the
XRF was located on the southeastern tip of a morphological irregular
source, the probable host of XRF\,011030. Following the methodology in
\citet{bkd02}, we measure the offset of the XRF to be 322$\pm$59 mas
East, 106$\pm$65 mas South (offset distance $r = 339 \pm 60$ mas) of
the apparent host center.  No spectroscopic redshift is known for this
galaxy. However, the offsets corresponds to 2.9 kpc in projection at a
redshift of unity (using $H_0 = 65$ km s$^{-1}$ Mpc$^{-1}$, $\Omega_m
= 0.3$, $\Omega_\Lambda = 0.7$); since angular diameter distance
is relatively insensitive to redshift over the range $z =$ 0.5--5,
this physical offset is expected to be accurate to $\sim$30\,\%.

\subsection{XRF\,020427}

While one non-transient X-ray source falls on the STIS field, the
nominal positions using the STIS and CXO headers did not coincide with
any obvious STIS counterpart. An investigation of the guide star
observation files, {\tt \_jif}, reveals that the STIS absolute
pointing determined using two Guide Stars in the {\it Faint Guidance
Sensors} (FGS) camera was suspect\footnotemark\footnotetext{The
predicted and the observed guide star offset differed by 1.678 arcsec,
suggesting the absolute pointing and roll angle were systematically
incorrect. The culprit was likely the primary guide star (GSC
0912300701 = TYC 2433221), which we calculated from the Tycho-2
catalog had moved 748 mas East, 1148 mas South of the GSC-listed
position at the epoch of the HST observation.  The approximate
magnitude and position angle of the absolute pointing offset was
confirmed by observations of WFPC2 parallel images (P.I.~S.~Casertano;
GO/PAR \#9318) taken concurrently with the STIS images and eight
sources in the USNO A2.0 astrometric catalog. The offset of the
Tycho-2 Guide Star from its GSC-listed position is approximately the
same as the shift of the native HST WCS relative to the ICRF (see
text).}.

Instead of absolute astrometry, we registered both the HST and CXO
frames independently to the LCO I-band image. We first found a world
coordinate system for the LCO image using 298
GSC2.2\footnotemark\footnotetext{{\tt
http://www-gsss.stsci.edu/gsc/gsc2/GSC2home.htm}} stars in the
field. To tie the STIS WCS to the LCO image, we first re-sampled the
STIS image by a factor of eight to 200 mas pixels. We then smoothed
that image to one arcsec seeing and then re-sampled the LCO image to
the same WCS zeropoint using IRAF/WREGISTER. Using IMALIGN and the 5
objects in common to both images, we then found a systematic shift of
(1.041 $\pm$ 0.051) arcsec East, (1.320 $\pm$ 0.051) arcsec North
between the native STIS and LCO WCSs.

To perform the CXO tie, we identified three stars in the LCO image
with counterparts in the CXO data.  Two of these stars are GSC 2.2
stars that were used to attach a WCS to the LCO image; for these
purposes, however, we use their positions as derived from the image
itself.  The third tie object is the Tycho star TYC 9123-1224-1; this
star was saturated in the LCO image and was not used to establish the
WCS; for CXO astrometry we use the Tycho catalog position (taking into
account the proper motion).  We derive a CXO coordinate shift of $40
\pm 130$ mas East, $230 \pm 120$ mas North from these three tie
objects.  Since the new WCS for the STIS image is derived from the LCO
image, our position for XRF~020427 within the ICRF frame is thus
determined to be $\alpha$(J2000) = 22\fh09\fm28\fs.2230,
$\delta$(J2000) = $-$65$^\circ$19\arcmin32\arcsec.031, with a
positional uncertainty of 280~mas (2 $\sigma$).

We measure the offset of XRF\,020427 to be 79$\pm$144 mas West,
42$\pm$142 mas South ($r = 90 \pm 145$ mas) of the apparent host
center; that is, the source position is consistent with the center of
the galaxy.  The offset corresponds to (0.78 $\pm$ 1.25) kpc in
projection at a redshift of unity. The localization of the XRF within
the host is consistent with, but more accurate than, the results from
the same data presented in \citet{frb+02}.

\section{Properties of the Host Galaxies}

The HST photometry of the hosts of both XRF\,011030 and XRF\,020427
was performed in an aperture of radius 25 drizzled pixels (635
mas). This aperture was selected as a trade off to include as much of
the galaxy light without inheriting large errors from an uncertain sky
background level. In the NIRC $K_s$-band image of the field of
XRF\,011030, we estimated the upper-limit of the host detection by
determining the r.m.s.\ background scatter in a 1.2 arcsec aperture
and used an aperture correction determined from a bright unsaturated
star.

The host flux of XRF\,020427 was found using IRAF/PHOT, with the
background level determined from randomly placed apertures in the
vicinity of the host.  The point spread function of the bright star
5.82 arcsec to the northeast of the host of XRF\,011030 cast a faint,
but noticeable, increase in the background level around the host
galaxy. To remove the contribution of the background to our aperture
photometry, we generated 45 independent apertures at the same radial
distance from the bright star and computed the total flux in each
aperture. We then subtracted the median of the ensemble of these
fluxes (after sigma-clipping to remove those apertures with
contaminating sources) and estimated the error on the background level
by taking the standard deviation of ensemble values.

After a determination of counts per second from the hosts, we
converted the instrumental flux to ST magnitudes using the photometry
header keywords in the images. We converted the Clear/Longpass colors
to the Johnson/Cousins system using template spectra from
\citet{bmp00} redshifted to $z=0-4$. The conversion to the $R$ magnitude
is fairly independent of the assumed template and redshift, and the
additional uncertainty introduced by the transformation is $\ale 0.1$
magnitudes. The transformation to $B-R$ color has a larger uncertainty
as it is more sensitive to the assumed template and redshift. The
errors in Table 1 include this systematic uncertainty. We note that
the ST colors of XRF\,011030, {\em uncorrected} for extinction, are
comparable to those derived from the bluest templates of
\citeauthor{bmp00} (a star burst spectrum). 
Therefore the reported $B-R$ should be considered strictly an
upper-limit to the true colors. The Galactic extinction toward the
fields are $E(B-V)$(XRF\,020427) = 0.029 mag and $E(B-V)$(XRF\,011030)
= 0.393 mag \citep{sfd98}.

The determinations of magnitudes of the XRF hosts differ from results
reported in the literature.  \citet{ctgsf02} reported $R = 23.3 \pm
0.2$\,mag and $B = 23.8 \pm 0.4$\,mag for the host XRF\,020427 based
upon groundbased imaging, brighter by $\sim$1 magnitude than reported
herein. However, \citet{ctgsf02} apparently photometered the entire
galaxy complex (within 4\arcsec $\times$ 2\arcsec\ region) which was
unresolved from the ground, but resolved into three distinct
components in HST imaging. We therefore believe that \citet{ctgsf02}
overestimated the host brightness by inclusion of the other nearby
galaxies in the aperture. The extinction-corrected magnitude of the
host of XRF\,011030 reported by \citet{fpk+02} is $V \approx 25$
mag. With $V-R=0.4\, (B-R)$ and $B-R \approx 0.5$\,mag we find $V
\approx 24.3$\,mag, a difference of 0.7\,mag. The details of the photometry
were not given by
\citet{fpk+02}, but the difference could be accounted for by choice of
aperture size and the method of determining the background (in the
sense that \citealt{fpk+02} overestimated the background relative to
our value). As stated, the presence of the nearby star casts a
non-negligible gradient of background light across the host. Both JSB
and PGvD photometred the host of XRF,011030 independently, using
different background estimators, and found consistent results for the
host photometry.

The extinction corrected $R$-band magnitudes of the XRF hosts give
some indication of the probable redshifts. Using the observed GRB host
luminosity function \citep{dfk+03} as an estimator and assuming that
XRFs are drawn from the same population of hosts, the redshift median
(10\%-ile, 90\%-ile) of hosts with the same $R$-band magnitudes of the
two XRF hosts is $z = 1.2$ (0.6, 2.6).  Interestingly, the color of
the host of XRF\,020427 cannot be matched by our template galaxy
spectra for $z \ale 1$. Given no significant drop in flux in the Clear
filter from Lyman $\alpha$ absorption, the colors of the host of
XRF\,020427 suggest that the source originated between a redshift of
order unity and $z \approx 3.5$. All of these redshift constraints are
relatively insensitive to the unknown intrinsic host galaxy spectral
energy distributions.  Even with these redshift constraints, we cannot
estimate the total energy output of the XRFs as the XRF fluence
measurements have yet to be reported.

Visually, the host of XRF\,011030 appears as a faint irregular galaxy
without a strong central condensation. Because of the overall low
signal--to--noise of the object, it is unclear whether a faint feature
$\sim$0.64 arcsec southeast from the center is part of the galaxy or
an unrelated source. Due to less foreground extinction, the host of
XRF\,020427 appears brighter than the host of XRF\,011030. The
XRF\,020427 host is asymmetric, which may be the result of a merger or
tidal interaction

We fitted two-dimensional exponential and S\'ersic morphological
profiles to the STIS/Clear images and found that the simplistic
profiles are inadequate to fully describe the complex morphologies of
the faint hosts.  The images are simply not deep enough to properly
characterize the shape of the galaxy profiles at large radii. Indeed,
as expected, the resulting modeled half-light radii ($r_h$) depend
rather strongly on the assumed profile. Simply taking the average
$r_h$ from our fits and using half of the full range for the error
gives $r_h$(XRF\,011030) = $560^{+230}_{-140}$ mas and
$r_h$(XRF\,020427) = $300^{+80}_{-70}$ mas. These values correspond to
physical sizes of $\sim$2--4 kpc, typical for bulges in the local
universe \citep{dj96}.

Irrespective of the assumed profile, the ellipticity of the galaxies
is fairly well-constrained: \\ $e$(XRF\,011030) = $0.46 \pm 0.04$ and
$e$(XRF\,020427) = $0.72 \pm 0.02$. The position angle (East of North)
of the semi-major axes is 32$^\circ \pm 2^\circ$ for XRF\,020427 and
$-$44$^\circ \pm 4^\circ$ for XRF\,011030. As seen in Figure 1, there
are two distorted galaxies at comparable magnitudes and colors within
a few arcseconds and to the south of the host of XRF\,020427,
suggestive of a tight grouping of physically related galaxies. No such
group is seen in XRF\,011030\footnotemark\footnotetext{There are
several faint blue galaxies in the field (that are also undetected in
$K_s$-band) at comparable magnitude to the XRF host (e.g., at located
at, RA: 20:43:33.49, DEC: +77:17:29.8; 20:43:35.76 +77:17:32.7;
J2000). Notably, one such blue galaxy (20:43:35.836, +77:17:21.34) has
an arc-like distortion around a red, possibly early-type galaxy, and
may therefore be lensed.}.

\section{Discussion}

To date, four XRFs (011030, 020427, 020903, 030723: \citealt{fkc+03})
have been followed up reasonably rapidly.  In all cases, long-lived
lower energy emission, i.e.~afterglow emission, appear to have been
discovered resulting in sub-arcsecond localizations.  For the first
two XRFs, we evaluate the probability of chance coincidence of the
afterglow positions and a random, unrelated galaxy.  Using the
offsets, host magnitudes, and the formulation in \citet{bkd02}, we
estimate this chance to be $P_{\rm ch} ({\rm XRF\,011030}) = 0.00797$
and $P_{\rm ch} ({\rm XRF\,020427}) = 0.00595$. Therefore, we believe
that, like most other GRBs localized to date, these XRFs are likely to
be physically associated with galaxies.

Accepting that these XRFs did indeed occur within their assigned
hosts, the photometric evidence presented herein suggests that at
least two members of the XRF class cannot be due to GRBs at high
redshift. This is consistent with the suggested lack of apparent time
dilation between XRFs and GRBs in their respective time histories
\citep{lr02}. In addition, the faintness of the hosts (as well as 
the photometry) suggests both sources occurred with $z \age 0.6$. This
poses difficultly for the original incarnation of the off-axis GRB
hypothesis for XRFs \citep{yin02}, which require $z \ale 0.4$ to be
bright enough for detection. Higher maximum XRF redshifts are possible
with narrowly beamed jets \citep{yin03a}, but the GRB collimation
angles required ($\ale 1^\circ$) do not appear consistent with the
inferred distribution of opening angles \citep{fks+01}.

We can compare the properties of the putative XRF host galaxies to the
host galaxies of GRBs. Adopting the half-light radii as found above,
the offsets amount to host-normalized projected offsets of
0.605$\pm$0.236 and 0.300 $\pm$ 0.490 (XRFs\,011030, 020427,
respectively) [see \citet{bkd02} for a formulation]. Relative to the
20 GRBs in the \citeauthor{bkd02} sample, the XRFs fall in the 33- and
23-percentile (XRFs 011030, 020427, respectively) in host normalized
offset. Since both XRFs are not located at the centers of their
respective hosts (as might be expected from an active nucleus origin),
from a large-scale perspective, this is tentative evidence that the
progenitors of at least some XRFs are related to stellar birthsites.

With an apparent irregular host of XRF\,011030 and with the host of
XRF\,020427 as a possible merger system, morphologically, the XRF
hosts are consistent with the diverse sample of GRB hosts
\citepeg{bkd02}. The galaxies associated with GRBs 990123
\citep{bod+99,ftm+99} and GRB 980613 \citep{htna+02} are examples of
disturbed GRB hosts. Photometrically, these XRF hosts appear somewhat
brighter than, but not significantly different from, the magnitudes of
GRB hosts \citepeg{hf99,dkb+01,pia01,sfc+01}.

While the general consensus is that GRB hosts are a blue, vigorously
star-forming population \citepeg{ldm+03}, no strong evidence to date
has been published to indicate that they are a population that is
significantly different from faint blue galaxies \citepeg{bdk01}. As
blue sources, the hosts of both of these XRFs fit this general trend
but until a more complete study has been published on GRB host colors
we cannot quantify the extent to which the XRF hosts standout (or fit
in). Although the $z \approx 1$ galaxy population has not yet been
fully characterized, indications are that a large fraction of galaxies
are blue and distorted. Therefore, the XRF hosts may be considered as
fairly typical in comparison to the field.

As with GRBs, one would need a larger sample, on the order of a few
dozen, to empirically demonstrate any significant link between XRFs
and star-forming galaxies. However, the first two well localized host
galaxies appear to be similar to those of GRB host galaxies and the
prima facie evidence suggest that XRFs, like GRBs, are intimately
related to star formation and subsequent stellar death.

{\it Note:} After this paper was submitted, \citet{fhg+03} placed an
upper limit to the redshift of XRF\,030723 of $z \approx 2.1$, from
the absence of strong Lyman $\alpha$ absorption in the spectrum of the
optical afterglow. This confirmed our claim that at least some XRFs
do not appear to be a manifestation of GRBs at very high redshift.

\acknowledgements

We thank the referee for insightful comments on the original version
of the manuscript.  We thank D.~Kaplan for assistance with the NIR
data reductions. J.S.B.\ is supported by a Junior Fellowship to the
Harvard Society of Fellows and by a generous research grant from the
Harvard-Smithsonian Center for Astrophysics.  S.~R.~K.\ and S.~G.~D.\
thank the NSF for support. Support for Proposal number
HST-GO-09180.01-A was provided by NASA through a grant from Space
Telescope Science Institute, which is operated by the Association of
Universities for Research in Astronomy, Incorporated, under NASA
Contract NAS5-26555. The authors wish to extend special thanks to
those of Hawaiian ancestry on whose sacred mountain we are privileged
to be guests. Without their generous hospitality, some of the
observations presented herein would not have been possible.


\begin{deluxetable}{lccccr}
\singlespace
\tablecolumns{7}
\tablewidth{0in}
\tablecaption{Photometry of the Hosts of XRF\,011030 and XRF\,020427\label{tab:hostphot}}
\tabletypesize{\small}
\tablehead{
\colhead{Source} & \multicolumn{4}{c}{Magnitudes\tablenotemark{a}} \\
\colhead{} & \colhead{Clear} & \colhead{Clear $-$ Longpass} &
             \colhead{$R_c$} & \colhead{$B - R_c$} & \colhead{$K_s$}}
\startdata
XRF\,011030 \ldots & $25.24 \pm 0.15$ & $0.30 \pm 0.29$ &
                     $24.11 \pm 0.18$ & $0.6 \pm 0.3$ & $>$21.70\tablenotemark{b} \\

XRF\,020427 \ldots & $24.38 \pm 0.05$ & $0.20 \pm 0.08$ &
                     $24.23 \pm 0.07$ & $0.50 \pm 0.15$
\enddata

\tablenotetext{a}{The STIS magnitudes are given uncorrected for Galactic 
extinction. The Johnson magnitudes and errors (cols.~4--5) are given
using galaxy templates to match the flux in the STIS filter (see
text); these magnitudes have been corrected for Galactic extinction
but assume no error in the Galactic extinction value from
\citet{sfd98}. No correction to the $B-R$ color of XRF\,011030 was applied as
it is quite dependent on the assumed spectral type and redshift; we
note, however, that the {\em uncorrected} colors are already as blue
as the bluest (star burst) template of \citet{bmp00}. The $B-R$ color
of XRF\,011030 should therefore be considered a strict upper
limit. All errors include systematic uncertainties in centroiding and
sky subtraction, but assume no error in the STIS zeropoints.}

\tablenotetext{b}{Reported is the
2-$\sigma$ non-detection limit for the XRF host.}

\end{deluxetable}

\end{document}